\documentclass{article}
\usepackage[utf8]{inputenc}
\usepackage{amsmath}
\usepackage{amssymb}
\usepackage{amsthm}
\usepackage{graphicx}
\usepackage{subcaption}
\usepackage{listings}
\usepackage{subcaption}
\usepackage{booktabs}
\usepackage[square,sort,comma,numbers]{natbib}
\usepackage{float}

\usepackage[a4paper, left=25mm, right=25mm, top=30mm, bottom=50mm]{geometry}
\usepackage{bbm}
\usepackage[dvipsnames]{xcolor}
\usepackage{footnote}
\usepackage{comment}

\usepackage{algorithm}
\usepackage{algpseudocode}

\usepackage{footnote}
\usepackage{ulem}
\usepackage{graphicx}
\usepackage{subcaption}
\usepackage{listings}
\usepackage{hyperref}
\usepackage{animate}
\usepackage{geometry}
\usepackage{bbm}
\usepackage{babel}
\usepackage{authblk}
\newcommand{\xx}{\mathbf{x}}
\newcommand{\yy}{\mathbf{y}}
\newcommand{\ee}{\mathbf{e}}
\newcommand{\zz}{\mathbf{z}}

\usepackage{setspace}
\begin{document}

\title{A biased random walk approach for modeling the collective chemotaxis of neural crest cells}

\author[1,2]{Viktoria Freingruber}

\author[3]{Kevin J. Painter}

\author[1]{Mariya Ptashnyk}

\author[4,2]{Linus Schumacher}

\affil[1]{The Maxwell Institute for Mathematical Sciences, Department of Mathematics, Heriot-Watt University, Edinburgh, EH14 4AS, UK}

\affil[2]{The Maxwell Institute for Mathematical Sciences, School of Mathematics, University of Edinburgh, Edinburgh, EH9 3FD, UK}

\affil[3]{Dipartimento Interateneo di Scienze, Progetto e Politiche del Territorio (DIST), Politecnico di Torino, Viale Pier Andrea Mattioli, 39, Turin, 10125, Italy}

\affil[4]{Centre for Regenerative Medicine, Institute for Regeneration and Repair, University of Edinburgh, 5 Little Frande Drive, Edinburgh, EH164UU, UK}

\date{}

\maketitle

\abstract{Collective cell migration is a multicellular phenomenon that arises in various biological contexts, including cancer and embryo development. `Collectiveness' can be promoted by cell-cell interactions such as co-attraction and contact inhibition of locomotion. These mechanisms act on cell polarity, pivotal for directed cell motility, through influencing the intracellular dynamics of small GTPases such as \textit{Rac1}. To model these dynamics we introduce a biased random walk model, where the bias depends on the internal state of \textit{Rac1}, and the \textit{Rac1} state is influenced by cell-cell interactions and chemoattractive cues. In an extensive simulation study we demonstrate and explain the scope and applicability of the introduced model in various scenarios. The use of a biased random walk model allows for the derivation of a corresponding partial differential equation for the cell density while still maintaining a certain level of intracellular detail from the individual based setting. }




\section{Introduction}\label{sec:intro}
Collective behaviour is ubiquitous in the living world and can be observed on various length-scales in many different cells and organisms \cite{sumpter2010collective, deisboeck2009collective, friedl2004collective, li2013collective}. One example is collective cell migration, which is involved in wound-healing \cite{friedl2009collective}, cancer metastasis \cite{friedl2009collective, rorth2009collective, li2013collective} and embryonic development \cite{friedl2009collective, schumacher2019collective}. Notably, both metastatic cancer cells and neural crest cells undergo epithelium-to-mesenchyme transition \cite{theveneau2012neural, friedl2009collective}. Hence, studying one can also give insights into the other. 

During vertebrate embryogenesis, the coordinated migration of neural crest cells (NCCs) plays a crucial role \cite{theveneau2012neural, friedl2009collective}. NCCs are a multipotent cell population that undergoes epithelial-to-mesenchymal transition during embryonic development \cite{theveneau2012neural}. In their epithelial state, they reside in the neural tube, a precursor tissue of the brain and spinal cord. However, in their mesenchymal state, NCCs either migrate individually or collectively through the embryo, differentiate into various cell types, and contribute to the formation of a multitude of developing tissues and organs \cite{weston1970migration}. Successful migration of NCCs involves a diverse array of processes that vary both between species and between different types of NCCs, such as cranial, vagal, trunk, or sacral NCCs \cite{bronner2016neural}.

A combination of both in vivo and in vitro experiments have demonstrated that NC migration involves various contributing mechanisms \cite{carmona2011complement, boer2015fascin1, burns2002ovo, theveneau2010collective}. In in vitro studies with zebrafish and Xenopus, the presence of co-attraction, contact inhibition of locomotion (CIL), and spatial confinement were sufficient for the successful collective migration of NCCs \cite{carmona2011complement}. Conversely, in vivo investigations indicate that a chemoattractant, such as \textit{Sdf1}, is required for the guided migration of NCCs in a particular stage of embryonic development in zebrafish \cite{boer2015fascin1}. Moreover, alternative in vivo studies involving vagal NCC transplantation into the sacral neuroaxis suggest that a chemoattractive cue is dispensable for vagal and sacral NCCs in chick and quail embryos, as these NCCs were capable of colonising the hindgut by following pathways usually taken by sacral cells \cite{burns2002ovo}. The reader is referred to \cite{theveneau2011can, theveneau2012neural} for reviews on mechanisms involved in NCC migration.

Using mathematical modeling in this context to simulate biological experiments in silico can contribute to the support or explanation of hypotheses and enhance understanding of the  mechanisms involved. Previous approaches to model the collective behaviour of NCC comprise, for example, off-latice individual based approaches \cite{mclennan2012multiscale, mclennan2015neural, schumacher2019neural}, cellular Potts models \cite{landman2011building}, agent-based models \cite{schumacher2019neural, wynn2012computational, wynn2013follow, carmona2011complement, szabo2016vivo, martinson2023dynamic} and discrete models with polygonal cell structures \cite{merchant2018rho, merchant2020rho}. The proposed models differ not only in their approach but also in the level of biological detail taken into consideration as well as in the type of mechanisms involved in the collective motion.

In the latter mentioned models \cite{merchant2018rho, merchant2020rho}, cell-cell contacts mediate the \textit{Rac1} and \textit{RhoA} levels on the cell membrane, which in return influence the direction of movement. This modeling approach captures more details about intracellular processes involved in cell migration, however, the complexity of the model limits to a pure simulation study, as opposed to less involved models that can potentially admit analytical investigation. In the individual-based models, mentioned above, the cell-cell and cell-environment interactions directly affect the direction of movement without considering the intracellular \textit{Rac1}-\textit{RhoA} bridge. On the other hand, derivation techniques can be applied to acquire a corresponding continuous equation for the cell density. For an extensive review on modelling techniques for collective cell migration the reader is referred to \cite{camley2017physical, giniunaite2020modelling}.


The aim of this work is to combine the advantages of a simple random walk model with the consideration of intracellular processes. Hence, we introduce a biased random walk model where the bias depends on the intracellular \textit{Rac1} levels surrounding the cell midpoint. By using a random walk approach we maintain the possibility of deriving a corresponding partial differential equation for the cell density following standard methods, e.g. \cite{painter2018random, stevens1997aggregation}. In this paper we present a two-dimensional hybrid lattice model describing the continuous evolution of the intracellular \textit{Rac1} dynamics and the discrete-in-time migration of the cells biased by the internal differences in \textit{Rac1}. In Section~\ref{sec:model} the mathematical model is introduced, before the simulation setup is presented in Section~\ref{sec:sim}. The numerical simulation results  are presented in Section~\ref{sec:results} and  discussed in Section~\ref{sec:disc}. In certain cases a representation of this model can be derived in form of a system of partial differential equations and the main principle of the derivation is demonstrated in Section~\ref{sec:derivation_mainchapter}.

\subsection{Mechanisms involved on the molecular scale}

One of the main mechanisms regulating cell polarity and hence the direction of movement for a motile cell is the intracellular \textit{Rac1-RhoA} system \cite{mayor2016front}. At locations with higher \textit{Rac1} activity, actin polymerization is enhanced and the extension of lamellipodia and filopodia is promoted; in other words, protrusions are formed at the leading edge of the cell \cite{mayor2016front}. Areas with a higher concentration of active \textit{RhoA} form at the trailing edge of the cell, where it controls myosin phosphorylation and actomyosin contractions that pull the back end of the cell forward \cite{mayor2016front}. The resulting cell-polarity is pivotal for directed cell motility \cite{mayor2016front}.

Chemokines, such as Stromal cell-derived factor 1 (\textit{Sdf1}) or Vascular endothelial growth factor (\textit{VEGF}), have been found to act as chemoattractants for specific subpopulations of NCCs \cite{belmadani2005chemokine, braun2002xenopus, mclennan2010vascular, mclennan2015vegf}. Additionally, it has been shown that \textit{Sdf1} is capable of stabilizing and maintaining the polarity of NCCs induced by cell-cell contacts \cite{theveneau2010collective}.

It has been observed that most neural crest (NC) cell types tend to migrate collectively, rather than individually \cite{theveneau2011can, kulesa2010cranial}. Short-range chemotaxis, up the gradient of the auto-produced complement factor \textit{C3a}, has been suggested as a mechanism for maintaining cohesiveness within the NC cell cluster \cite{carmona2011complement}. Specifically, when \textit{C3a} binds to its receptor \textit{C3aR}, it activates the \textit{Rac1} pathway \cite{carmona2011complement}, inducing sufficient cell polarization in escaping cells to enable them to return to the group \cite{carmona2011complement}.

A counterbalance to the co-attraction is a short-ranged repulsion mechanism between colliding cells, referred to as contact inhibition of locomotion (CIL) \cite{mayor2010keeping}. This phenomenon, first observed and described in vitro in fibroblasts \cite{abercrombie1953observations,abercrombie1954observations} causes cells to alter their migration direction upon collision and move in opposing directions. In \textit{Xenopus} and zebrafish, NCCs have been shown to exhibit CIL in both in vivo and in vitro studies \cite{carmona2008contact}. Additionally, it has been demonstrated that the induction of cell polarity through cell-cell contact is pivotal for the collective chemotaxis in NCCs \cite{theveneau2010collective}. Mechanistically, on the site of cell-cell contact \textit{RhoA} increases \cite{carmona2008contact} while \textit{Rac1} activity decreases concurrently \cite{theveneau2010collective}. These mechanisms are induced by N-Cadherin and Wnt/PCP signalling processes \cite{carmona2008contact, theveneau2010integrating}.

Other important factors influencing NCC migration include surrounding spatial constraints or chemorepellents \cite{shellard2019integrating}. Extracellular matrix components, such as \textit{Versican} or signalling factors such as ephrins, semaphorins and DAN act as chemorepellents and prevent mixing between separate NCC stream and keep the NCC from invading certain tissue \cite{theveneau2012neural}. In \textit{Xenopus}, the repulsive signal \textit{Versican} is located at the border of the NC and by acting as spatial constraint it enhances NC migration \cite{szabo2016vivo}. 

The following articles provide detailed information on chemotaxis in NCC \cite{shellard2016chemotaxis}, chemical and mechanical signals involved in NCC migration \cite{shellard2019integrating}, contact inhibition of locomotion \cite{mayor2010keeping} and molecular mechanisms involved in regulation of \textit{Rac1} and \textit{RhoA} \cite{theveneau2010integrating}, respectively.


\section{Model description}\label{sec:model}

In this section, we present a novel hybrid model to describe the collective movement of neural crest cells (NCCs). Our model adopts a biased random walk in discrete time and space, where the internal state variables that govern the bias evolve continuously between two discrete jumps.
The distinctive feature of our approach lies in the unique cell shape that we consider. Previous models of collective migration and chemotaxis have employed random walks of point particles \cite{stevens1997aggregation}, which enable the derivation of a corresponding partial differential equation for the total cell density 
\cite{painter2018random}. However, this approach fails to capture any spatial heterogeneity with respect to an internal cell state. In contrast, other models such as those presented in \cite{merchant2018rho, merchant2020rho} employ a polygon shape to represent the cell and track internal cell states at multiple boundary locations as well as within the cell. While these models capture more of the complexity of internal cell processes, deriving an equation for the total cell density from these models using standard techniques is challenging, necessitating the use of large scale simulations to gain insight.

\begin{figure}
    \centering
    \includegraphics[width=0.75\textwidth]{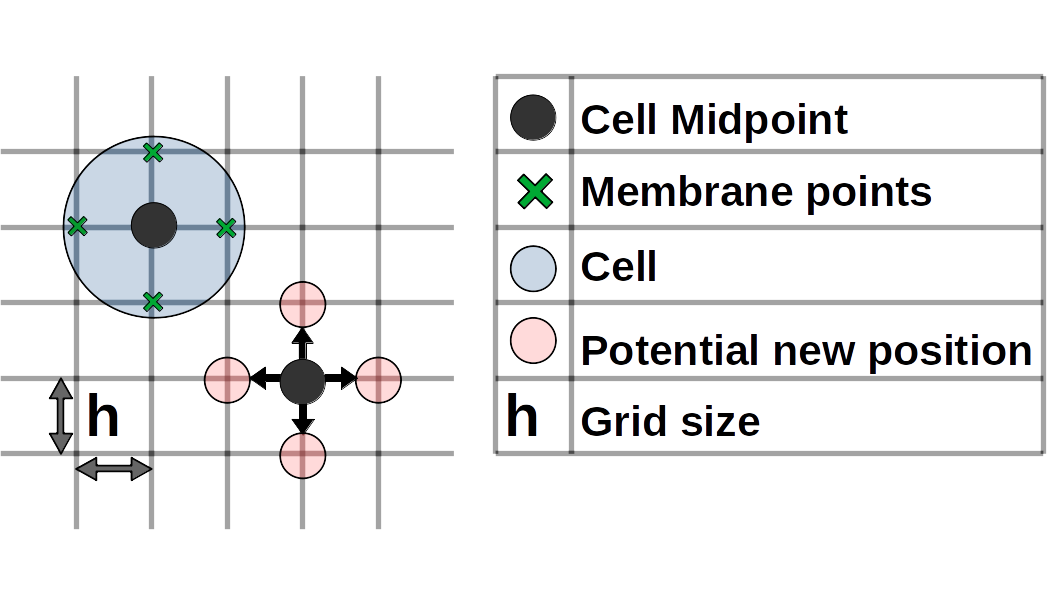}
    \caption{Visualization of a single cell on a two-dimensional lattice: Cell center (grey dot); membrane points (green crosses); Area occupied by a cell (blue circle); Potential locations for jumping (light red dots) }
    \label{fig:my_cell}
\end{figure}

In our approach we aim to combine the advantages of using point particle models and spatially extended models. To achieve this, we track the center of each cell, which acts as a point particle, and incorporate additional spatial detail through the implementation of a surrounding membrane composed of adjacent grid points for every cell (as depicted in Figure~\ref{fig:my_cell}). By doing so, we are able to retain information on the internal cell states, while still having the potential to derive continuous equations for the cell densities.

\begin{figure}
     \centering
     \begin{subfigure}[b]{0.49\textwidth}
         \centering
         \includegraphics[width=\textwidth]{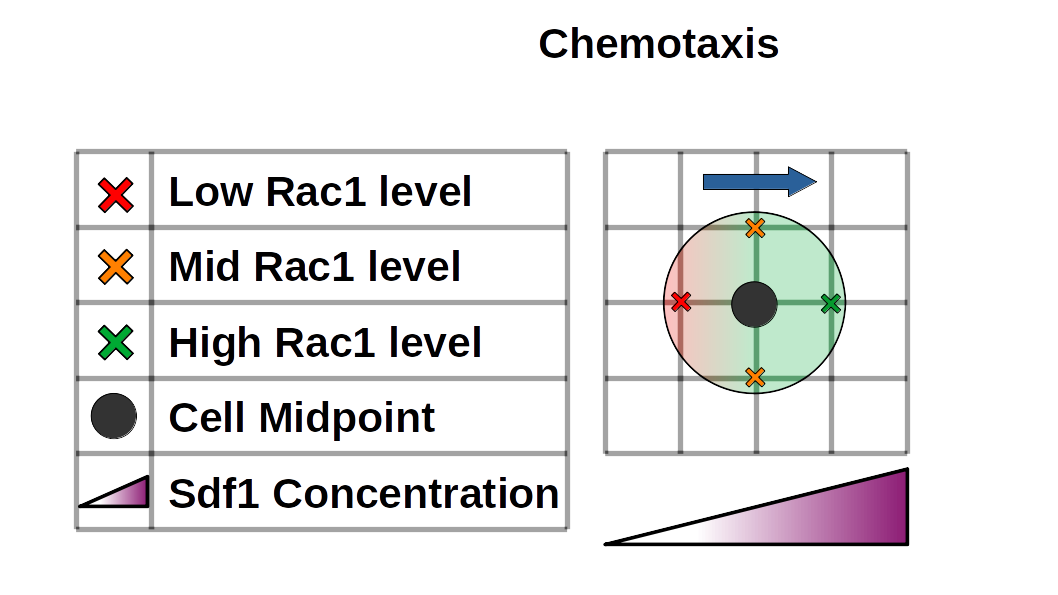}
         \caption{Chemoattraction}
         \label{fig:sdf1}
     \end{subfigure}
     \begin{subfigure}[b]{0.49\textwidth}
         \centering
         \includegraphics[width=\textwidth]{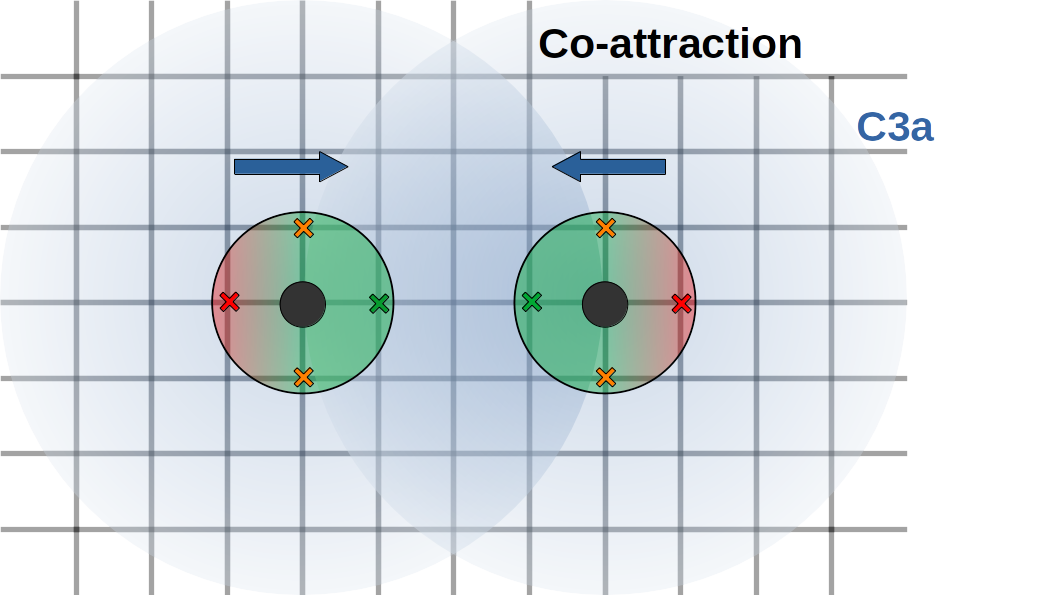}
         \caption{Co-attraction}
         \label{fig:coa}
     \end{subfigure} \\
     \hfill
     \begin{subfigure}[b]{0.49\textwidth}
         \centering
         \includegraphics[width=\textwidth]{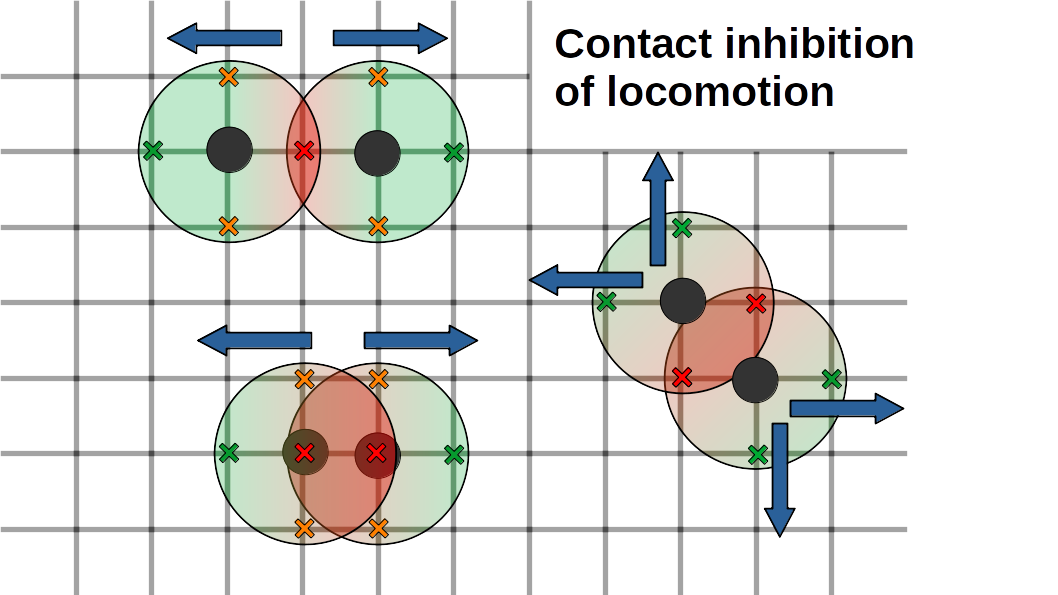}
         \caption{Contact inhibition of locomotion}
         \label{fig:cil}
     \end{subfigure}
          \begin{subfigure}[b]{0.49\textwidth}
         \centering
         \includegraphics[width=\textwidth]{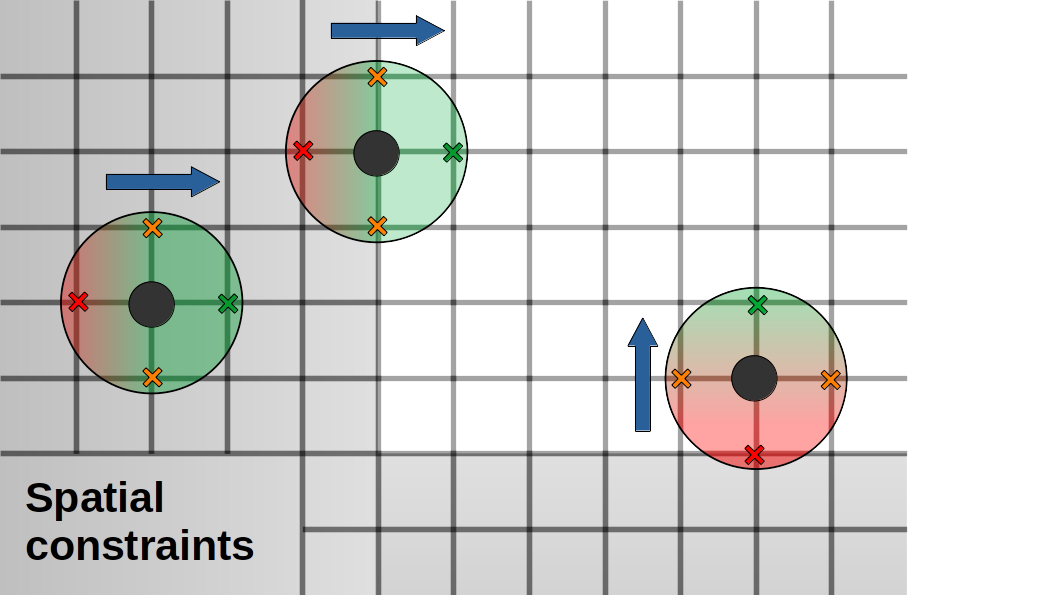}
         \caption{Spatial constraints}
         \label{fig:wall}
     \end{subfigure}
        \caption{Schematic of the four mechanisms incorporated in the model, each displayed individually in the absence of the others. Area occupied by one cell is displayed with a color gradient to emphasis the Rac1 polarization, Blue arrows indicate most probable direction of movement. Influences of a) Chemoattractant, b) Co-attractive C3a profiles (blue circles), c) Contact inhibition of locomotion and d) spatial constraints on the internal Rac1 states.}
        \label{fig:diagrams}
\end{figure}

The key model assumptions are as follows:
\begin{itemize}
\item \textit{Domain}: Our model assumes that the cells migrate on a two-dimensional discrete lattice, characterized by a fixed grid size. In certain simulations we will also impose spatial constraints to mimic the effects of surrounding tissue or chemorepellents. Specifically, the confinement is represented as a corridor with walls on three sides that are ``soft", allowing cells to cross them and exit the corridor. However, this action incurs a negative impact on the internal variable that drives the directional bias of the cell. Thus, cells are biased to migrate away from regions outside the corridor, to account for the constraints imposed by the surrounding tissue~\cite{szabo2016vivo}.

\item \textit{Number of cells:} A population of  $N$ cells is considered and cells are  placed on the grid with the  position of  their midpoints at one of the lattice points. We do not consider any cell division or cell death, therefore the number of cells remains constant.

\item \textit{Shape of cells:} Each cell is characterised by  the midpoint, representing the central cell position, and  four adjacent points, mimicking the cell membrane, see Figure~\ref{fig:my_cell}. We assume that the cells do not undergo any growth or division, thus the distance between the membrane point and the cell position is consistently equal to the grid size, and the shape of the cell remains constant.

\item \textit{Movement of cells:} Cells undergo a biased random walk with a fixed velocity on the lattice, meaning they move a distance  $h$ (grid size) per time step. The direction of movement is biased by the differences in the internal state variable along one axis, resulting in a tendency for the cells to move in directions with higher concentrations of the internal state. We allow the cells to overlap either partially or entirely, as our model does not account for all the spatial dimensions present in reality. Hence, the occurrence of cells with the same location should be interpreted as cells that are stacked on top of one another rather than cells occupying the same physical space. We do not allow for cell rotation in our model.

\item \textit{Internal state variable:} We consider the intracellular molecule \textit{Rac1} as the internal state variable that regulates cell motility. The support of \textit{Rac1} for each cell is limited to the directly adjacent grid points of the central cell position and changes over time. The directional bias is influenced by the difference of the internal state value along one axis, where a higher difference in \textit{Rac1} leads to a higher probability in moving in the direction where more \textit{Rac1} is present. The internal variable is subject to the following extracellular mechanisms:
\begin{itemize}
     \item Increase by chemoattractant \textit{Sdf1}: \\ We assume that the chemoattractant's profile is constant in time and independent of the cell population. Furthermore, the profile should be non-negative and continuous. We consider different profiles for different simulation scenarios.

    \item Increase by co-attraction molecule \textit{C3a}: \\
    The process of co-attraction between cells is governed by the chemoattractant molecule \textit{C3a}, which is synthesized and released by each cell. We assume that these processes occur at a faster time scale than those related to the polarization of the cell or cell movement. Accordingly, we consider each cell to have its own stationary \textit{C3a}-profile, which is radially symmetric and non-negative, and decreases with distance from the cell, e.g. as proposed by \cite{carmona2011complement}. Note that although the individual \textit{C3a} profile does not change in time due to any production, diffusion or degradation, the total \textit{C3a} profile changes in time due to the movement of the cells.
    We assume that an individual cell can sense \textit{C3a} released by neighbouring cells if the cells centers have a distance less than $R$, where \textit{C3a} surrounding different cells are treated in an additive way. In other words, the concentration of \textit{C3a}  around an individual cell is  positive only within a radius  $R$ around the cell midpoint.
     
    \item Decrease due to contact inhibition of motion: \\
    Upon contact with neighbouring cells, the local \textit{Rac1} concentration decreases at the point of contact. As ``contact" we consider the overlapping of one or more center or membrane points occupied by a cell, see~Figure~\ref{fig:cil}.

    \item Natural inactivation: \\ 
    We assume that \textit{Rac1} is naturally inactivated at a fixed rate.
    
    \item Decrease upon contact with spatial constraints: \\  The spatial confinement representing surrounding tissue is implemented as a chemorepellent that decreases the local \textit{Rac1} activity. Consequently, the cells are discouraged from leaving the corridor due to the negative impact of the confinement on their internal variable.

\end{itemize}
\end{itemize}

\subsection{Model formulation}

Let $\xx = (x_1,x_2) \in h \mathbb{Z} ^2$ be a point on the two-dimensional discrete lattice with grid size $h$ and let $\mathcal{A} (\xx) :=  \{ \xx \pm h \ee_1, \xx \pm h \ee_2 \} $  and $\mathcal{A}^0(\xx) := \{ \xx \} \cup \mathcal{A}( \xx )$ denote the set of all adjacent grid points to  $\xx$, excluding and including $\xx$, respectively. Here, $\ee _j$ denotes the $j$-th unit vector. As for the biological interpretation of these two sets, the former can be seen as the discretized membrane of the cell centered at $\xx$ and the latter is then the area occupied by the whole cell.

Consider a number of $N$ individual cells located on the lattice and let $P^i(t_n)$ denote the position of the $i$-th cell at time $t_n \in \tau \mathbb{N}$ with initial position $P^i(0) = \xx ^i_0$, for $i=1,...,N$. Furthermore, let $\mathcal{P}(t) = \{ P^i(t) : i=1,...,N\}$ be the set of all cell positions at time $t$. We assume that each cell leaves its position within a time interval $[t_n,t_{n+1})$ with probability 1.

The position at time $t_n$ of the $i$-th cell can be written as the sum of all jumps that were made up until that time,
\begin{equation}
\label{eq:discreteparticle}
P^i(t_n) = \xx^i_0 + \sum_{k=1}^n Y^i_k,
\end{equation}
where $Y^i_n \in \{ \pm h \ee _j\}_{j=1,2}$ denotes the outcome of the decision the $i$-th cell makes for the direction of movement in the time interval $[t_{n-1},t_n)$.

Let $C^i(\xx,t)$ denote the internal state variable of the $i$-th cell at $(\xx,t)$ for $t\geq 0$. Note, that in comparison to the evolution of a cell position discrete in time, the time variable here is continuous. Biologically, this variable represents the concentration of the intracellular molecule \textit{Rac1} at the cell membrane. The temporal change in this variable is twofold:
\begin{enumerate}
\item Every discrete time step, when a jump occurs, the cell midpoint and membrane points change.
Hence, the support of $C^i(\xx,t)$ changes every discrete time step. The support of $C^i(\xx,t)$, i.e.~the points where $C^i(\xx,t) \neq 0$, is restricted to the adjacent grid points to the central cell position, i.e. $\text{supp}\left( C^i(\xx,t) \right) = \mathcal{A}(P^i(t_k))$ for $t_k \leq t < t_{k+1}$.
\item Between two discrete jumps, i.e.~for $t_k \leq t < t_{k+1}$, the internal state will be subject to changes described by the following ordinary differential equation
\begin{equation}
\label{eq:rac}
\frac{d}{d t} C^i(\xx,t) = \begin{cases} 
\lambda_1 S^i_1 (\xx) + \lambda_2 S_2 ^i (\xx,t_k)- B(\xx,t_k) C^i (\xx,t)  & \text{ for } \xx \in \mathcal{A}(P^i(t_k)), \\
0 \quad & \text{ else }, 
\end{cases}
\end{equation}
where $B(\xx,t_k) =  \lambda_3 S_3 ^i (\xx,t_k) + \lambda_4 
+ \lambda_5 b(\xx)$ and $\lambda_1$, $\lambda_2$ represent the activation rate of \textit{Rac1} by a chemoattractive and a co-attractive cue, respectively, $\lambda_3$ and $\lambda_5$ are the deactivation rates of \textit{Rac1} by contact inhibition of locomotion and spatial constraints, respectively, $\lambda_4$ is the natural deactivation rate of \textit{Rac1}. The rates $\lambda_j$, for $j=1,\dots,5$, are positive constants. The system is completed by initial conditions 
 \begin{equation}
 C^i(\xx,t_k) = \begin{cases} C^i( \xx-Y^i_k, t_k) \quad & \text{for} \quad \xx \in \mathcal{A}(P^i(t_{k})), \\
0 \quad & \text{else},  \end{cases}
 \end{equation}
where $\xx - Y^i_k$ is the previous location of this membrane point. 
For $t=0$ we define
$$
C^i(\xx,0) = \begin{cases} 
C_0 > 0 \quad & \text{for} \quad \xx \in \mathcal{A}(\xx_0^i),\\
0 \quad &\text{else}.
\end{cases}
$$
In \eqref{eq:rac}, $S_1(\xx) \geq 0$ is the profile of the outer chemoattractant \textit{Sdf1} which we assume to be constant in time, $S_2^i (\xx,t_k)$ is the cumulative co-attractive profile for the $i$-th cell, sensing all the neighbours that are located within a radius $R$ of the $i$-th cell, i.e. $P^j(t_k) \in \mathcal{P}(t_k) \setminus P^i(t_k)$ such that $\| P^j(t_k) - \xx \|<R$ for $\xx \in \mathcal{A}(P^i(t_k))$ and $j=1, \ldots, N$. The individual co-attraction profiles are given by 
\begin{align}
\label{eq:S2}
     S_2 ^i (\xx,t_k) & = \sum_{\substack{P^j(t_k) \in \mathcal{P}(t_k) \setminus P^i (t_k) \\ \| P^j(t_k) - \xx \| < R}}  M \exp \left( - \frac{1}{w} \|  P^j(t_k) - \xx \| \right) ,
\end{align}
with parameters $M$ and $w$, 
    and $S^i_3(\xx,t_k)$ acts as a counter of overlapping cells
    $$
    S_3^i (\xx,t_k) =  \sum_{\substack{j \leq N_{\text{cells}} \\ j \neq i}} \mathbbm{1}_{\mathcal{A}^0(P^j(t_k))}(\xx). 
    $$
    Finally, $b(\xx)$ describes potential spatial obstacles or constraints. We have that $b(\xx)=0$ wherever cells should be able to move freely
    and  $b(\xx) >0$ at locations of spatial constraints, e.g. where the walls of a corridor are located.
\end{enumerate}

Let $\mathcal{T}^i_{\xx \to \yy}$ denote the probability of the $i$-th cell jumping from position $\xx$ to position $\yy$. We assume that the cells move with constant speed $h/\tau$, which in particular means that remaining at the same position and jumps of size larger than $h$ are impossible and that a cell leaves its position with probability~$1$ within one discrete time step $\tau$. Formally this means 
$$\mathcal{T}^i_{\xx \to \yy}  = 0 \text{ for all } \yy \notin \mathcal{A}(\xx) \; \text{ and } \; \sum_{\yy \in \mathcal{A}(\xx)} \mathcal{T}^i_{\xx \to \yy} = 1. $$
The probability of the $i$-th cell jumping from a position $\xx$ at time $t_k$ to an adjacent position $ \yy $ at time $t_{k+1}$ will be assumed to have the form
$$
\mathcal{T}^i_{\xx \to \yy} = \frac{\mathcal{\hat{T}}^i_{\xx \to \yy}}{\sum_{\yy \in \mathcal{A}(\xx)} \mathcal{\hat{T}}^i_{\xx \to \yy}},
$$
with
$$
\begin{aligned}
\mathcal{\hat{T}}^i_{\xx \to \yy} =&  \mathcal{\hat{T}} \left(C^i(\yy,t_{k+1}) - C^i ( \xx - (\yy - \xx), t_{k+1}) \right) \\
=& \alpha + \beta g \left( C^i(\yy,t_{k+1}) - C^i ( \xx - (\yy - \xx), t_{k+1}) \right), 
\end{aligned}
$$
where $g$ is an increasing ($g'(\cdot) > 0 $) and bounded ($\lim_{\xx \to \pm \infty} g(\xx) = \pm c$) function and  positive constants $c, \alpha, \beta$,  such that  $\alpha> \beta c$ to ensure $\mathcal{T}^i_{\xx \to \yy} > 0$. The function $g$ represents the bias coming from the  internal  state variable of the cell. The boundedness assumption on $g$  ensures positivity of the probabilities and is biologically reasonable. We choose an increasing function to represent the assumption that a higher difference in the internal state variable leads to an increased polarity and with that an increased probability of moving into a beneficial direction. We will compare two bias functions $g(\cdot)$:
\begin{enumerate}
    \item \textit{Symmetric bias function:} We assume that $g(\cdot)$ is an odd function, i.e. that $g(- \xx) = - g( \xx)$. Due to the oddness of the function the positive bias in one direction equals the negative bias in the opposite direction. This leads to a normalization constant $4 \alpha$ in the two spatial dimensions,
    $$
\mathcal{T}^i_{\xx \to \yy} = \frac{\mathcal{\hat{T}}^i_{\xx \to \yy}}{4 \alpha }.
$$ 
Note that the probability of moving vertically or horizontally is equal to $1/2$ due to that choice. This also means that the maximal probability of jumping into any direction is $1/2$ in this case.
\item \textit{Asymmetric bias function:} In case a cell is confined on three of its sides, the symmetric bias function might yield unsuitably weighted probabilities since the probability of moving into two of those confined sides is still $1/2$. Hence, we also consider an asymmetric bias function that only adds to the probability if there is a positive benefit of moving into a certain direction, see~Figure~\ref{fig:ic2d} (left panel).
\end{enumerate}


\section{Simulation setup}\label{sec:sim}

\begin{figure}
    \centering\includegraphics[width=0.475\textwidth]{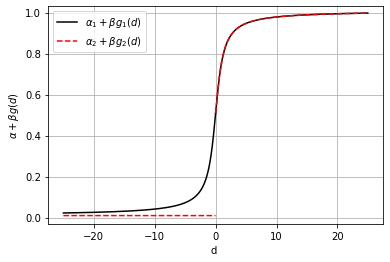}
\includegraphics[width=0.475\textwidth]{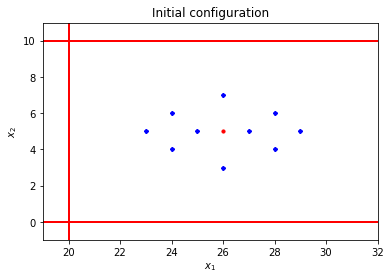}
    \caption{(Left) Different gain functions, $\alpha_1 = 0.6$, $\alpha_2 = 0.1$, $\beta_1 = \beta_2 = 1/\pi$. (Right) Initial non-dimensionalized configuration in two spacial dimensions. The blue dots mark the initial positions of the 10 cells, the mean position is marked with a red dot. The boundary of the corridor is depicted by red lines located at $x_1 = 20$, $x_2 = 0$, $x_2 = 10$. Note, that only some simulations were conducted in the domain with boundaries, in other cases there was no spatial constraint.}
    \label{fig:ic2d}
\end{figure}

The initial setup:
A number of $N_{\text{cells}} =10 $ cells are initially placed as in Figure~\ref{fig:ic2d} (right panel). 
The initial configuration aims to represent a densely organised cell cluster that has just undergone epithelial to mesenchymal transition, where cells are still maintaining cell-cell contacts at $t=0$ but will be able to move independently in the next instant. The pseudo-code can be found in Section~\ref{appx:Algorithm}.

We will differentiate between several main simulation set-ups:

\begin{itemize}
\item \textit{Two different bias functions $g(\cdot)$:}
\begin{enumerate}
    \item \textit{Symmetric bias function:} We use an odd function 
    $$ g_1(\cdot) = \arctan (\cdot) $$ 
    and positive constants $\alpha_1$, $\beta_1$ such that $\mathcal{\hat{T}}^i_{\xx \to \yy} = \alpha_1 + \beta_1 g_1(d) > 0 $, where $d =C^i(\yy,t_{k+1}) - C^i ( \xx - (\yy - \xx), t_{k+1}) $. The choice of this function inherently represents the assumption that there can be both, negative and positive, directional bias.
    \item \textit{Asymmetric bias function:} We use 
    \begin{equation*}
     g_2(d) =    \begin{cases} 0 \quad \text{for} \quad d \leq 0,\\
     \arctan (d) + \frac{\pi}{2} \quad \text{for} \quad d > 0,          
        \end{cases}
    \end{equation*}
    and positive constants $\alpha_2, \beta_2$ such that $\mathcal{\hat{T}}^i_{\xx \to \yy} = \alpha_2 + \beta_2 g_2 (d) >0$, see~Figure~\ref{fig:ic2d} (left panel).
\end{enumerate}
Note that we will chose a different value for $\alpha_1$ and $\alpha_2$ in both these cases, see~Table~\ref{tab:parameters}, as the value of $\alpha_1$ in the symmetric case has to be sufficiently large to ensure positivity of the probability.
\item \textit{With or without spatial confinement:} In cases with spatial confinement, the domain within which cells are free to move is a corridor displayed in Figure~\ref{fig:ic2d} (right panel). The area outside the  corridor, with boundary marked by red lines,  mimics the presence of chemorepellents or spatial constraints. Note that the cells are technically able to move across the boundary since the presence of corridor walls makes it only less likely but not impossible for them to move there. For these simulations we will use the following corridor function $b(\xx)$:

\begin{align*}
    b(\xx) = \begin{cases} -x_1 +20, \quad \text{for} \quad x_1<20, \\
    x_2 - 10, \quad \text{for} \quad x_2 >10, \\
    -x_2, \quad \text{for} \quad x_2 <0,
    \end{cases} 
\end{align*}
in equation~\eqref{eq:rac}, where it is made increasingly difficult to move further away from the corridor with the presence of a chemorepelling substance such as \textit{Versican} in the surrounding tissue increases with distance to the corridor. This assumption is not based on any biological observations and  purely practical,  ensuring that the cells move back into the corridor with higher probability.

 \item \textit{With or without co-attraction, contact inhibition of locomotion and natural inactivation of \textit{Rac1}:} 
In cases without one or more of these mechanisms, the corresponding rate in the ODE (\ref{eq:rac}) is set to $0$. We distinguish between different cases with different combinations for (in-)active mechanisms, see~Table~\ref{tab:casesresults}. As a benchmark scenario we will consider the case without CIL, co-attraction, natural inactivation and spatial confinement, i.e.~the chemoattractant is the only active component for these simulations. This case can be regarded as the chemotaxis of individual cells without any cell-cell interactions.

\item \textit{With different chemotaxis settings:}
    \begin{enumerate}
    \item Without a chemoattractive profile.
    
    \item With a constant-in time chemoattractive profile 
    $$ S_1^i(\xx) = f_1(\xx), \; \text{ with } \; f_1(\xx)=\frac{x_1+100}{100}.$$
    
    \item With a chemoattractive profile that can only be sensed by a cell if the cell is already sufficiently polarized due to cell-cell interactions, i.e. if sufficient \textit{Rac1} is present \cite{theveneau2010collective}. This switch is modelled with a Hill function that is dependent on the local \textit{Rac1} concentration
    $$ S^i_1(\xx) = f_2(\xx), \;  \text{ where } f_2(\xx)= \frac{C^i(\xx,t)^n}{C^i(\xx,t)^n+K^n} f_1(\xx),$$ 
    where $K$ is the threshold for sufficient cell-cell interaction to sense the chemoattractant \textit{Sdf1} and $n\geq 1$.
    
    \end{enumerate}
 
 \item \textit{A strong and weak chemotaxis regime:} This will be done by a modulation of parameter $\lambda_1$ in equation~\eqref{eq:rac}. 

\end{itemize}

\begin{table}[ht]
\begin{center}
\begin{minipage}{\textwidth}
\begin{center}
\caption{Table of  cases considered in numerical simulations: co-attraction (COA), contact inhibition of locomotion (CIL), natural inactivation (NI), spatial constraints (SC)}    \label{tab:casesresults}
    \begin{tabular}{@{}l|c|c|c|c|c@{}}
    \toprule
        Case &  COA & CIL  & NI & SC \\ \hline \hline
       Benchmark (BM) & - & - & - & -  \\ \hline
    A & - & - & - & + \\ \hline
    B  & - & - & + & + \\ \hline
    C  & - & + & + & + \\ \hline
    D  & + & - & + & + \\ \hline
    E  & + & + & + & + \\ \hline
    F  & + & + & - & + \\
       
    \end{tabular}
    \end{center}
\end{minipage}
\end{center}
\end{table}

\begin{table}[ht]
\begin{center}
\begin{minipage}{\textwidth}
\centering
\caption{Dimensionless parameters} \label{tab:parameters}
\begin{tabular}{@{}llll@{}}
\toprule
Parameter & Description & Value & Source \\
\midrule
 $h$ & Step size & $1 \hat{=} 20 \mu m$ & \cite{woods2014directional}\\
 $\tau$ & Time increment & $1 \hat{=} 7 \min$ & \cite{woods2014directional} \\
 $\alpha_1$ & Randomness coefficient & 0.6 & - \\
  $\alpha_2$ & Randomness coefficient & 0.1 & - \\
 $\beta_1 = \beta_2$ & Weight of the bias & $1/ \pi$ & - \\
 $M$ & Maximal intensity of co-attraction & 32 & \cite{merchant2018rho} \\
 $d$ & Width of co-attraction profile & 8h & \cite{woods2014directional} \\
 $R$ & Radius of Co-attraction & $5h$ & \cite{merchant2018rho} \footnotemark[1] \\
 $\lambda_1$ & Rac1 activation rate by Sdf1 & $ \tilde{\lambda} \cdot 8 \cdot 10^{-3}$ & \cite{merchant2018rho}\\
 $\lambda_2 $ & Rac1 activation rate by C3a & $\tilde{\lambda} \cdot  2.4 \cdot 10^{-4}$ & \cite{merchant2018rho} \\
 $\lambda_3$ & Rac deactivation rate by Cil & $\tilde{\lambda} \cdot 8 \cdot 10^{-3} $ &  \cite{merchant2018rho} \\
 $\lambda_4$ & Rac1 base inactivation rate & $\tilde{\lambda} \cdot 2 \cdot 10^{-4}$ & \cite{merchant2018rho} \\ 
  $\lambda_4$ & Rac1 inactivation rate by spatial constraints& $\tilde{\lambda} \cdot 2 \cdot 10^{-1}$ & \cite{merchant2018rho} \\ 
  $\tilde{\lambda}$ & Parameter used for non-dimensionalization & $ 4 \cdot 10^{2} s$ & \cite{merchant2018rho} \\
\end{tabular}
\end{minipage}
\end{center}
\end{table}

\subsection{Parametrization of the mathematical model}
\label{sec:param}
We follow \cite{woods2014directional} and assume that cells have a fixed diameter of $40 \mu m$ and choose the grid size accordingly so that $h=1 \equiv 20 \mu m$. 
NCC migrate with an average speed of approximately $3 \mu m / \min$, on the faster end in a solitary state and slower when they are part of a cluster \cite{woods2014directional}. For this reason, we chose the time step $\tau = 1 \equiv 7 \min$ so that we obtain a constant speed of $20 \mu m / 7 \min \approx 2.86 \mu m / \min$. The system is non-dimensionalized in time and space in a way such that $h=1, \tau = 1$. The rates $\lambda_{i}$, $i=1,\dots,5$, are non-dimensionalized with the non-dimensionalization parameter $\tilde{\lambda} = 4 \cdot 10^2 s \approx 7 \cdot 60 s$. In reality cells slow down upon contact with other cells and then take some time to regain their regular speed. In this model we do not consider any variation in speed, for simplicity. 

The corridors, through which  NCCs migrate, vary in width depending on the exact type of the NCC as well as on the species \cite{szabo2016vivo}. Generally, larger clusters migrate through wider tubes \cite{szabo2016vivo}. We follow \cite{woods2014directional} and chose the width of the corridor to be $200~\mu m$, which corresponds to 5 cell diameters. 
The number of cells in a group of interacting NCCs also depends on the specific NCC type and species, for computational reasons we simulated a cluster comprised of~$10$~cells.

For the co-attractive profile we assume a steady state distribution at every time step since the diffusion of \textit{C3a} acts on much faster timescales than the cell migration. According to \cite{woods2014directional} this steady state distribution can be described by a Bessel function, and approximated by a decaying exponential with an approximate half maximum length of $110~\mu m$.  Adapting  to our model, this leads to the choice of 
$d = 8$ in equation~\eqref{eq:S2}. The maximum co-attractive strength  and the rates of \textit{Rac1} activation and deactivation were adapted from~\cite{merchant2018rho}.

\subsection{Statistics for numerical simulation results}
To describe the two-dimensional simulation results, each simulation setup was ran $N_{\text{sim}}=100$ times. To quantify the migrated distance we use the mean and standard deviation of the mean $x_1$ position
$$
M_{x_1} (t_k) := \frac{1}{N_{\text{sim}}} \frac{1}{N_{\text{cells}}} \sum_{j \leq N_{\text{sim}}}  \sum_{i \leq N_{\text{cells}}} x^i_{1,j},
$$
evaluated at a given time $t$. 
To quantify the dispersion of clusters we will use the mean root-mean-squared-deviation of the clusters
$$
\text{MRMSD} (t_k) := \frac{1}{N_{\text{sim}}} \sum_{j \leq N_{\text{sim}}} \sqrt{ \frac{1}{N_{\text{cells}}} \sum_{i \leq N_{\text{cells}}} \| \Bar{\xx}_j(t_k) - P^i_j(t_k) \| },
$$
where $P^i_j(t_k)$ is the position of the $i$-th cell in the $j$-th simulation at time $t_k$, $\Bar{\xx}_j(t_k)$ is the mean position of the cluster in the $j$-th simulation at time $t_k$ and $N_{\text{cells}}$ is the number of cells used for one simulation. While other measures can be conceived, we utilise these for their relative simplicity. In the Figure~\ref{fig:summaryplot} the simulation results are summarized by plotting the measure for migrated distance $M_{x_1}(t) - M_{x_1}(0)$ against the measure for dispersal $\text{MRMSD}(t) - \text{MRMSD}(0)$.

\section{Results}\label{sec:results}

\begin{figure}[t!]
\centering
\includegraphics[width=0.75\textwidth]{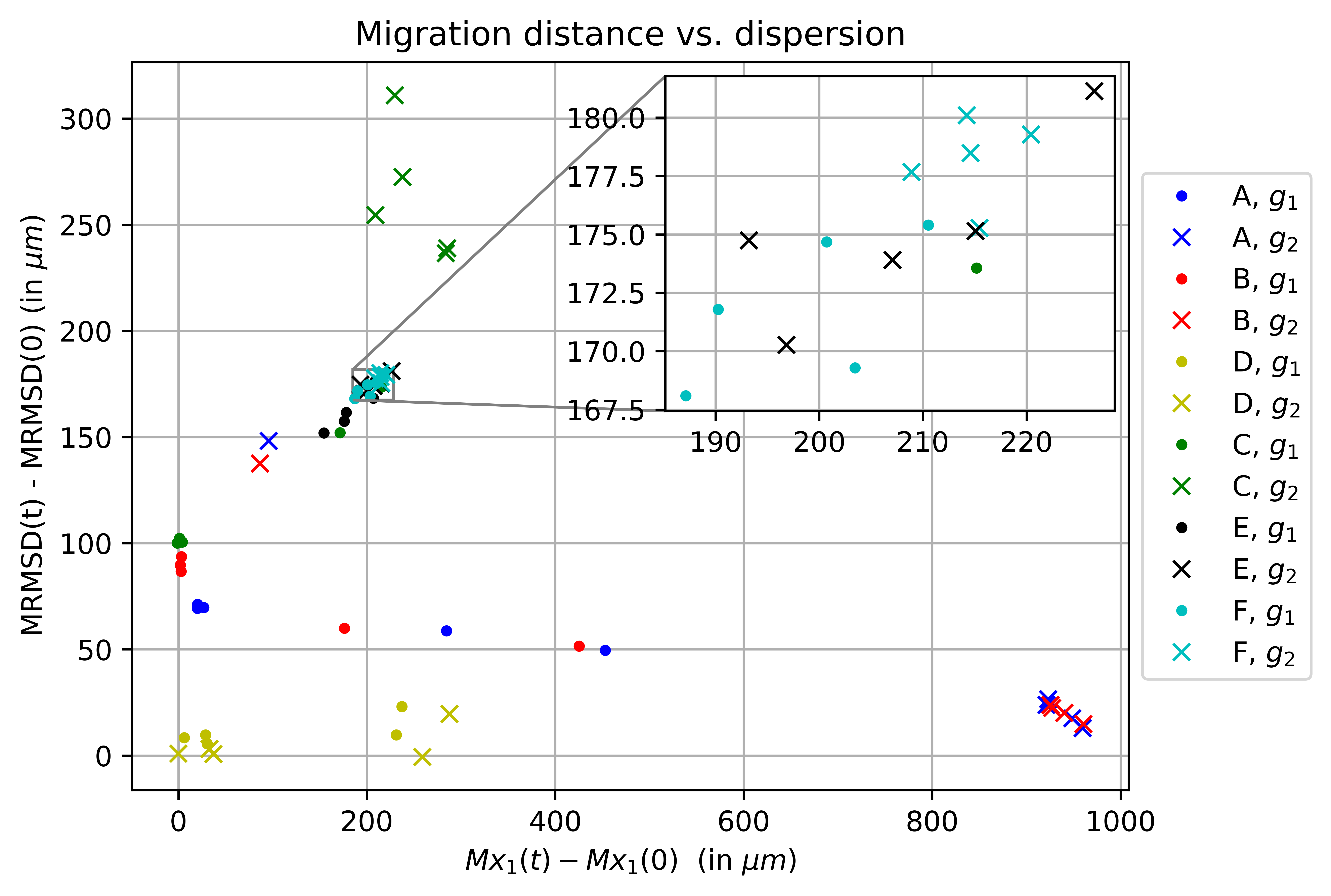}
\caption{Summary of simulation results: migration distance $M_{x_1}(350\min) - M_{x_1}(0)$ vs. dispersion $\text{MRMSD}(350 \min) - \text{MRMSD}(0)$ of the clusters.}
\label{fig:summaryplot}
\end{figure}

\begin{figure}[tp!]
    \centering
    \begin{subfigure}[b]{0.8\textwidth}
    \includegraphics[width=1\textwidth]{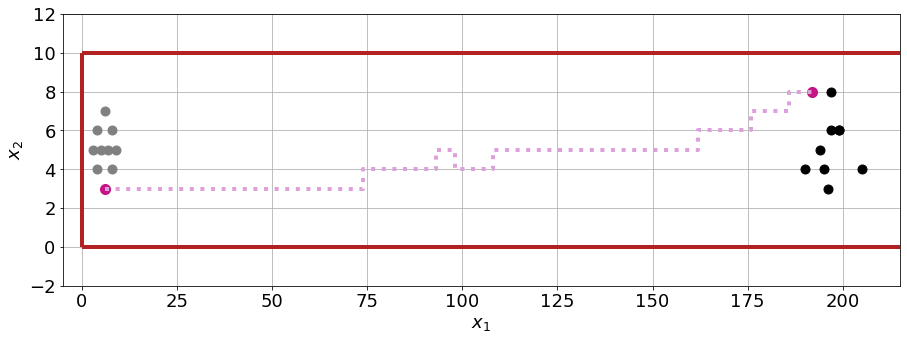}
    \caption{Individual simulation without cell-cell interactions, large~$\lambda_1$}
    \label{fig:res:OIIAg2hl}
        \end{subfigure} \\
         \begin{subfigure}[b]{0.8\textwidth}
    \includegraphics[width=1\textwidth]{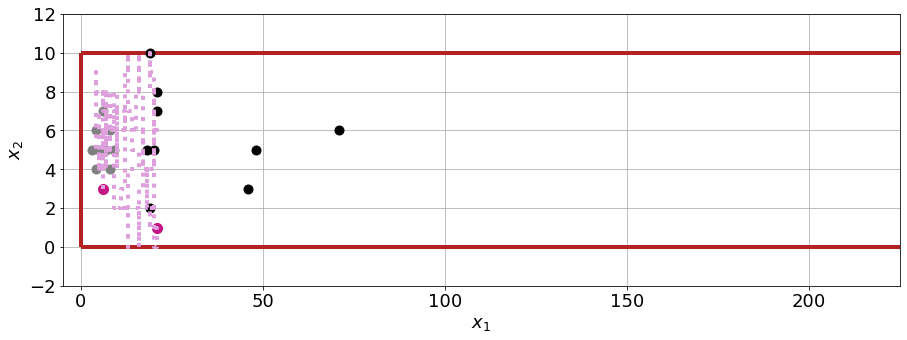}
    \caption{Individual simulation with inactive CIL, active co-attraction, small~$\lambda_1$}
    \label{fig:res:IIA3_g2_ll}
        \end{subfigure} \\   
                 \begin{subfigure}[b]{0.8\textwidth}
    \includegraphics[width=1\textwidth]{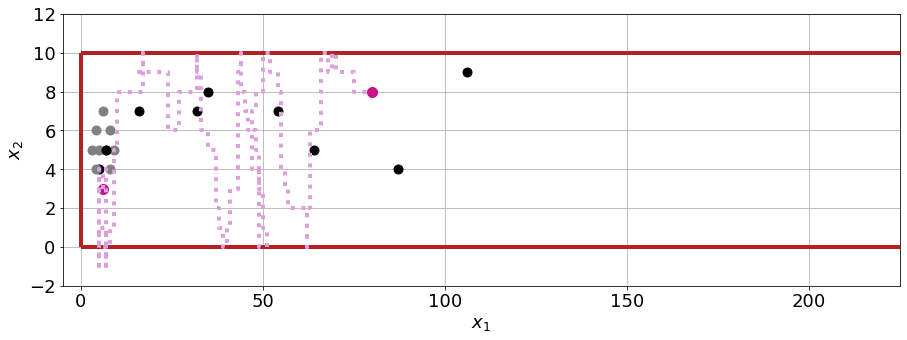}
    \caption{Individual simulation with active CIL, active co-attraction, small~$\lambda_1$}
    \label{fig:res:IIA1_g2_ll}
        \end{subfigure} \\   
        \caption{Individual realizations of simulations with $S_1(\xx) = f_1(\xx)$, spatial confinement and gain function $g_2$. Initial conditions and positions at $t = 350 \min$ are displayed by grey and black dots, respectively. The trajectory of an individual cell is visualized by a purple line. Note that the dimensions are stretched in the vertical direction, which assists visualisation of individual cell positions and tracks.}
        \label{fig:res:indsims}
\end{figure}

\begin{figure}[tp!]
    \centering
    \includegraphics[scale = 0.65]{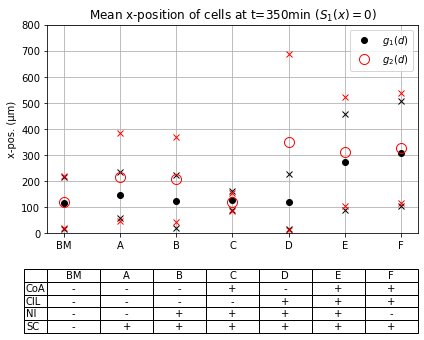}
    \caption{Mean $x_1$-position at $t = 350~\min$ in different scenarios with or without co-attraction (CoA), contact inhibition of locomotion (CIL), natural \textit{Rac1} inactivation (NI) and spatial constraints (SC) in the absence of a chemoattractive substance. }
    \label{fig:res:S1_0}
\end{figure}

\begin{figure}[tp!]
    \centering
    \begin{subfigure}[b]{0.65\textwidth}
    \includegraphics[width=1\textwidth]{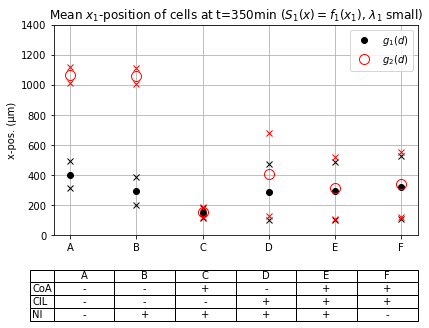}
    \caption{}
    \label{fig:res:lowlambda_a}
        \end{subfigure} \\
         \begin{subfigure}[b]{0.65\textwidth}
    \includegraphics[width=1\textwidth]{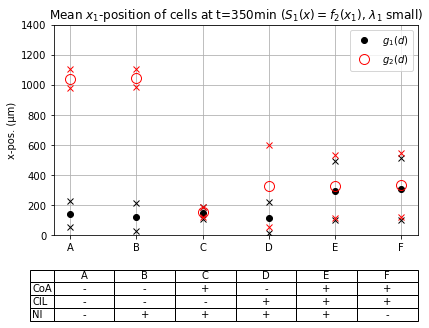}
    \caption{}
    \label{fig:res:lowlambda_b}
        \end{subfigure} \\   
        \caption{Mean $x_1$-position at $t = 350~\min$ with chemoattractive profile $S_1(x) = f_1(x)$ and small $\lambda_1$ in different scenarios with or without co-attraction (CoA), contact inhibition of locomotion (CIL), natural \textit{Rac1} inactivation (NI) and spatial constraints (SC) }
        \label{fig:res:lowlambda}
\end{figure}

\begin{figure}
    \centering
    \begin{subfigure}[b]{0.65\textwidth}
    \centering
    \includegraphics[width=1\textwidth]{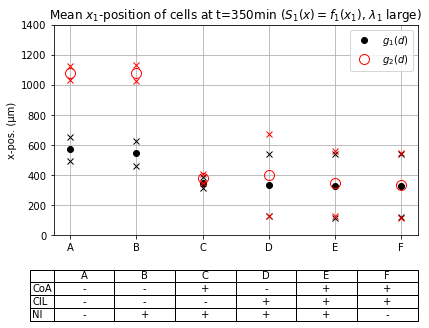}
    \caption{ }
    \label{fig:res:largelambda_a}
        \end{subfigure} \\
         \begin{subfigure}[b]{0.65\textwidth}
         \centering
    \includegraphics[width=1\textwidth]{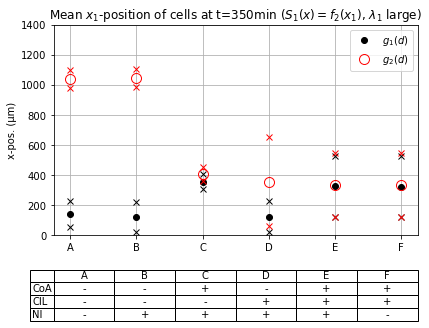}
    \caption{ }
    \label{fig:res:largelambda_b}
        \end{subfigure}
\caption{Mean $x_1$-position at $t = 350~\min$ with chemoattractive profile $S_1(x) = f_1(x)$ and large $\lambda_1$ in different scenarios with or without co-attraction (CoA), contact inhibition of locomotion (CIL), natural \textit{Rac1} inactivation (NI) and spatial constraints (SC)}
\label{fig:res:largelambda}
\end{figure}

We provide a summary figure of the simulation results in Figure~\ref{fig:summaryplot}, where the output of each simulation is represented in terms of $(\mbox{migrated distance},\mbox{cluster dispersal})$ after 350 minutes; points positioned further to the right indicate higher migrated distances, while points positioned higher up the axis indicate greater dispersal of the initial cluster. Simulation scenarios A-F are described in Table~\ref{tab:casesresults}, and $g_1$ and $g_2$ are the two different bias functions used. Figure~\ref{fig:res:indsims} illustrates typical model output for some representative simulations. Figures~\ref{fig:res:S1_0}-\ref{fig:res:largelambda} provide more detailed statistics of certain scenarios. In Figures \ref{fig:Dispersionovertime_S1_0},\ref{fig:Dispersionovertime_S1_f1} and \ref{fig:Dispersionovertime_S1_f2} we display the temporal evolution of the key measures presented in the summary plot Figure~\ref{fig:summaryplot}. More detailed descriptions follow below.

\subsection{Spatial confinement promotes collective cell migration}

All simulations were conducted with and without spatial constraints, with the results consistently indicating that confinement facilitates directed migration. The spatial constraints were implemented as triggering a decrease in   \textit{Rac1}, i.e.~as an inhibition of cell movement factor. In the absence of spatial constraints or a dominant chemoattractive cue, the initially dense cell clusters dispersed in the majority of cases. Therefore, spatial constraints appear to be not only advantageous but also necessary for maintaining a certain degree of cohesiveness within the migrating cluster, considering the other assumptions we made on the system, especially the assumptions made on the co-attractive strength, see~Figure~\ref{fig:res:S1_0}. 

\subsection{Asymmetric gain function allows for a strong cell polarity and increases migration distance}

We conducted simulations with two different bias functions, $g_1$ and $g_2$, to investigate different abilities to maintain stable cell polarizations. When considering $g_1$, the polarization alternates on average every other time step between pointing in a horizontal or vertical direction, whereas $g_2$ allows for a strong polarization, potentially guiding the cell down the most favorable path. Based on these assumptions, we expected better guided movement with $g_2$ as the probability of moving in the most advantageous direction can exceed $1/2$.\\
This expectation was confirmed in all simulation experiments, as shown in~Figures~\ref{fig:res:S1_0}, \ref{fig:res:lowlambda}, and~\ref{fig:res:largelambda}, where cell clusters migrated further when $g_2$ was used, sometimes by a considerable margin. 
The most crucial distinction between using an asymmetric and symmetric gain function is evident in cases `A' and `B', where interactive mechanisms are inactive, and the chemotactic cue is the only factor influencing the bias. The asymmetric function allows the bias to dominate the probability of moving in the direction of the chemotactic cue, see~Figure~\ref{fig:res:OIIAg2hl}, while the usage of $g_1$ leads to less directed movement, see~Figures~\ref{fig:res:lowlambda} and~\ref{fig:res:largelambda}.

\subsection{In the absence of a chemoattractant}
Prior studies have presented in silico findings regarding experiments that exclude chemoattractive cues, solely focusing on the impact of cell-cell interactions and potential spatial restrictions \cite{carmona2011complement, szabo2016vivo}. Our present study employs analogous experiments to assess the consistency of our model with regards to these circumstances.

In the absence of all cell-cell and cell-environment interaction mechanisms, cells move randomly without any preferential direction, and hence the unnormalized rate of movement in either direction is equal to $\alpha_i + \beta_i g_i(0)$, resulting in a normalized probability of $1/4$ for all cells. This scenario serves as the base case (BM), and the results remain the same for both gain functions, as shown in Figure~\ref{fig:res:S1_0}. Mechanisms are added one at a time to demonstrate its influence. The addition of spatial constraints promotes cell migration down the corridor, as demonstrated in case `A' in Figure~\ref{fig:res:S1_0}. The natural inactivation rate of \textit{Rac1} reduces the `memory time'  of previous interactions, causing cells to migrate slightly less far when natural inactivation (NI) is active, see case `B' in Figure~\ref{fig:res:S1_0}. The inclusion of co-attraction encourages cells to cluster densely, preventing collective or individual migration, see~`C' in Figure~\ref{fig:res:S1_0}. However, the addition of CIL promotes the dispersal of individual cells while still maintaining a certain level of cohesiveness within the cluster, see case `E' in Figure~\ref{fig:res:S1_0}. Omitting the natural inactivation of \textit{Rac1} slightly increases the migrated distance, see case `F' in Figure~\ref{fig:res:S1_0}, which is similar to the effect observed in case `B' compared to case `A'. When the deactivation rate is set to zero, the \textit{Rac1} level retains information regarding previous encounters with neighboring cells or the environment. All of the above observations hold true for both $g_1$ and $g_2$. When the co-attraction mechanism is inactive, the system behavior becomes strongly dependent on the form of gain function used. The gain function $g_2$ allows for strong singular cell polarity, and consequently, cells actively disperse, resulting in an increased migration distance on average but with a large standard deviation. In contrast, the symmetric gain function $g_1$ restricts cell polarity and hence results in less drastic dispersal. In summary, in the absence of a chemoattractant cells migrate farthest  when both co-attraction and contact inhibition of locomotion are active.

\subsection{Low rate of \textit{Rac1} up-regulation by chemoattractant \textit{Sdf1}}

Unsurprisingly, introduction of a chemoattractive cue leads to enhanced migration along the corridor compared to the absence of a chemoattractant, see~Figure~\ref{fig:res:lowlambda}. Note  that the chemoattractive profile is constant in the second spatial dimension and, hence, any movement in the second spatial dimension is due to other mechanisms or randomness; we note further that the vertical axis scale differs between Figures~\ref{fig:res:lowlambda} and~\ref{fig:res:S1_0}.
As before, an absence of CIL while co-attraction is active results in high clustering of cells but with the prevention of efficient migration, see case `C' in Figure~\ref{fig:res:lowlambda}, and Figure~\ref{fig:res:IIA3_g2_ll}. This observation does not depend on the gain function or chemoattractive profile used. 

As previously mentioned, the gain function $g_2$ generates a strong cell polarity and, hence, the bias leading the cells up the gradient of the chemical cue is comparably strong in the absence of other competing interaction mechanisms, see~cases `A' and `B' in Figure~\ref{fig:res:lowlambda_a}. This observation holds even when the \textit{Rac1} activation rate by the chemoattractive profile is dependent on the active \textit{Rac1} concentration itself, and the \textit{Rac1} activating co-attraction is switched off, see  `A' and `B' in~Figure~\ref{fig:res:lowlambda_b}. In these cases almost every movement is up the chemical gradient, similar to the case in Figure~\ref{fig:res:OIIAg2hl}.

The gain function $g_1$ does not create such a drastic cell polarization and, hence, the average migration down the corridor in the absence of any interaction mechanisms other than the chemical cue is not as efficient here,  
see `A' and `B' in Figure~\ref{fig:res:lowlambda_a}.
For both gain functions $g_1$ and $g_2$, adding the co-attraction adds a competing mechanism to the chemical cue and as a result the clusters migrate less far, see cases~`E' and `F' in Figure~\ref{fig:res:lowlambda_a}.

The most interesting result in this setting is obtained from using the gain function $g_1$ and the Hill-type response to the chemoattractive profile. Here, the cells perform best when both the co-attraction and co-repulsion mechanism are active, see~cases~`E' and `F' in~Figure~\ref{fig:res:lowlambda_b}. The both cases `E' and `F' perform similarly, although the clusters move marginally further if there is no natural inactivation. An example of one realization of case `E' is shown in~Figure~\ref{fig:res:IIA1_g2_ll}.

\subsection{Higher rate of \textit{Rac1} up-regulation by chemoattractant \textit{Sdf1}}

When the \textit{Rac1} activation rate by the chemoattractant is increased further, the migration distance increases significantly in case `C' but only marginally in  cases `E' and `F', see~Figure~\ref{fig:res:largelambda_a}. In cases `A' and `B' the migratory distance did  increase only when $g_1$ was used as it was already at its maximum in the $g_2$ case, see~Figure~\ref{fig:res:largelambda_a}. In  case `C' co-attraction and strong chemotaxis together manage  to move the tight cluster further down the corridor, see~Figures~\ref{fig:res:largelambda_a} and~\ref{fig:res:largelambda_b}.

When the Hill-type response to the chemoattractant, i.e.~$S_1(\xx) = f_2(\xx)$, and the symmetric gain function $g_1$ are used, the combined effect of CIL and co-attraction causes again a stronger migration,  compared to cases `A' and `B', but is now similar to when co-attraction only was used, see~Figure~\ref{fig:res:largelambda_b}.


\section{Derivation of approximating PDE system} \label{sec:derivation_mainchapter}

An objective when formulating \eqref{eq:discreteparticle} was to strike a balance between complexity and simplicity: i.e. allow each cell to be governed by a ``tractable level'' of intracellular heterogeneity. To demonstrate this, we provide a proof of concept analysis in which we derive the governing continuous PDE system for the evolving cluster distribution. The subsequent exploration and analysis of the derived system is left for future study.

While we are interested in finding the position of the $i$-th cell at time $t$, for a continuous description of the process it is natural to consider the probability distribution of the $i$-th cell's location at time $t$.
In a completely unbiased setting in two dimensions, the probability of jumping in either of the four directions would be $1/{4}$ and in the continuous limit we would obtain a pure diffusion equation for the time evolution of the cells. In our setting we assume that the probability of jumping in a direction is biased by an internal state variable $C^i(\xx ,t)$ that is non-zero on the cell membrane for ${\xx} \in \mathcal{A} (P^i(t))$.

Let  $p^i(\xx,t)$ denote the probability of the $i$-th cell being located at position $\xx$ at time $t\geq 0$. Initially, we have $p^i(\xx_0^i,0) = 1$ and $p^i(\xx,0) = 0 $ for all $\xx \neq \xx_0^i$. The change in probability of the $i$-th cell being located at $\xx$ at time $t+\tau$ is then
\begin{equation}\label{eq:mastereq_singlecell}
p^i(\xx ,t+\tau) - p^i(\xx ,t) = \sum_{\yy \in \mathcal{A}(\xx )} \mathcal{T}^i_{\yy \to \xx} p^i(\yy,t) - \mathcal{T}^i_{\xx \to \yy} p^i(\xx,t),
\end{equation}
where $\mathcal{T}^i_{\xx \to \yy}$ is the probability of the $i$-th cell jumping from its current position $\xx$ to an adjacent position $\yy \in \mathcal{A}(\xx)$ within a time interval $[t,t+\tau)$, as defined in Section~\ref{sec:model}. In \cite{stevens1997aggregation} different dependencies of the transition rates on the bias inducing variable are discussed and corresponding PDEs are derived. 

\begin{figure}
    \centering
    \includegraphics[width= 0.8\textwidth]{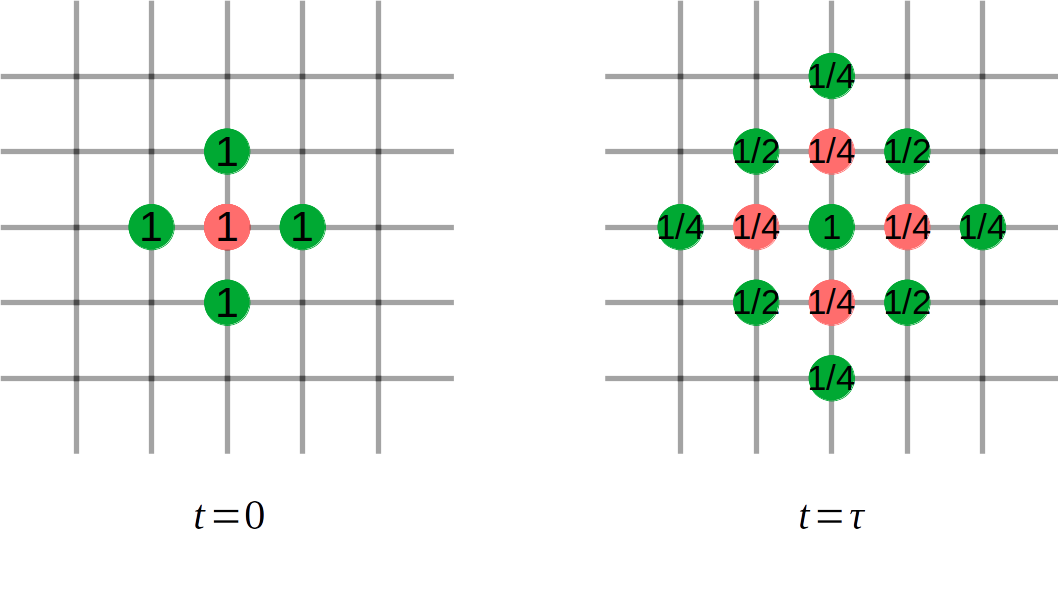}
    \captionof{figure}{Left: Initial position of cell midpoint (red) and membrane points (green) with associated probabilities. Right: After one step in a purely random walk, cell midpoint positions (red) and membrane points (green) with associated probabilities.}
\label{fig:probability}
\end{figure}

When following the individual trajectory,  the position of the centre of a cell and the associated membrane points is clearly defined by  $P^i(t)$  and $\mathcal{A}(P^i(t))$. In the continuous description,  while the support of the  probability $p^i(\xx,t)$ will initially  be just a point, the support of both the centre of the cell and the membrane points will disperse, see Figure~\ref{fig:probability}. In Figure~\ref{fig:probability} the probability distributions of a single cell (red dots) and its membrane (green dots) are displayed at times $t=0$ and $t= \tau$. The green point in the middle of the right hand side of Figure~\ref{fig:probability} holds membrane information associated with all the four adjacent potential cell locations. Therefore, in order for the master equation formulation to make sense we have to specify the cell midpoint and  membrane points associated with it.

Let  $H^i_\yy (P^i(t),t) := C^i(\yy,t)$     
denote the value of the internal state variable at $(\yy ,t)$ associated with the midpoint $P^i(t)$ of the cell. Then 
$$
\mathcal{T}^i_{\xx \to \yy} = \mathcal{T} \left( H^i_y(\xx,t) - H^i_{\xx-(\yy-\xx)}(\xx,t) \right),
$$
where the first argument of $H^i$ is the cell centre and the subscript $\yy$ is the considered membrane point, and the distance $\| \xx - \yy \|=h$.
Assuming that $\tau$ and $h$ are small, using  in \eqref{eq:mastereq_singlecell} the Taylor series expansion about $(\xx,t)$, and taking the limit as $h,\tau \to 0$ such that $\lim_{h,\tau \to 0} ({h^2}/{\tau}) =: D$ is positive and finite, yields
\begin{equation}\label{PDE_1}
\begin{aligned}
\frac{\partial}{\partial t } p^i(\xx,t) & =D \nabla \cdot \left[ \mathcal{T}(0) \nabla p^i(\xx,t) - 4 \mathcal{T}'(0) p^i(\xx,t) \nabla H^i(\xx,t)  \right], \\
p^i(\xx,0) &= \delta (\xx - \xx^i_0).
\end{aligned}
\end{equation}
Note, that assumption $\lim_{h,\tau \to 0} ({h^2}/{\tau}) < \infty$ implies ${h}/{\tau} \to \infty$, hence the effective speed is infinitely large and the solution $p^i(\xx,t)$ of \eqref{PDE_1} is non-zero at positions that are very far away from the starting position after a very short amount of time. Hence, the PDE is not necessarily a good approximation of the biased random walk for small $t$ unless one ignores very small densities.

The master equation for internal state variable at the membrane point $\yy$ corresponding to the cell centered  at $\xx$ is
$$
\begin{aligned}
&H^i_\yy (\xx,t+\tau) - H^i_\yy(\xx,t) = F(S_1(\xx),S_2^i(\xx,t),S_3^i(\xx),H^i_\yy(\xx,t))\\
&+\sum_{\zz \in \mathcal{A}(\xx)} \mathcal{T}_{\zz \to \xx}(H^i_\xx(\zz,t) - H^i_{2\zz-\xx} (\zz,t)) H^i_{\yy - (\xx-\zz) }(\zz,t) - H^i_\yy(\xx,t),
\end{aligned}
$$
where $F$ represents the change of \textit{Rac1} in a time interval $[t,t+\tau)$ defined by the mechanisms described in Section~\ref{sec:model}.
The Taylor expansion around $(\xx,t)$, combined  with the limit as $h,\tau \to 0$, yields 
$$
\begin{aligned}
   \frac{\partial}{ \partial t} H^i (\xx,t)
   =& 
 \lambda_1 S_1^i (\xx) + \lambda_2 \tilde{S}_2 ^i (\xx,t) - \left(  \lambda_3 \tilde{S}_3 ^i (\xx,t) + \lambda_4 + \lambda_5 b(\xx) \right) H^i (\xx,t) \\
  & + D \nabla \cdot \Big[ \mathcal{T}(0) \nabla H^i (\xx,t)  - 4 \mathcal{T}'(0)  H^i(\xx,t)  \nabla H^i(\xx,t) \Big]
 , \\
   H^i(\xx,0) =&  C_0 \delta_0(\xx^i_0 - \xx),
\end{aligned}
$$
where  $\tilde{S}_2^i(\xx,t)$ and $\tilde{S}_3^i (\xx,t)$ are smooth interpolations of the functions $S_2^i (\xx,t)$ and $S_3^i(\xx,t)$. 

Equation \eqref{PDE_1}, together with the equation for $H^i$, provides the continuous description for the dynamics of the  probability $p^i(\xx,t)$ for a cell to be at position $\xx$ at time $t$, when initially located at $\xx_0^i$.  The interactions between cells in a domain are included in  $\tilde{S}_2^i$ and $\tilde{S}_3^i$ and the impact of the environment is modeled by $S_1^i$. 

While this represents a complicated and coupled system of equations, analysis and numerical simulations for continuous equations may still be more efficient than the discrete model. This lies out of scope of the present article, where the primary aim was an analysis of the underlying random walk model and to provide a `proof of concept' for the continuous derivation. For further comment, we refer to the discussion.

\section{Discussion}\label{sec:disc}

In this paper we have formulated a random walk model for collective cell movement dynamics, motivated by instances of NCC migration during embryonic development. NCCs achieve migration in a variety of ways -- from individuals to coordinated clusters, according to cell type and species -- and a wide variety of guiding factors have been proposed \cite{szabo2018}: extracellular chemoattractant gradients, ligand-mediated co-attraction between cells, contact-induced cell-to-cell repulsion (contact inhibition of locomotion), and confinement that results from inhibiting cues in the surrounding extracellular matrix. Motivated by populations of cranial {\it Xenopus} NCC cells that migrate as a collective group \cite{theveneau2010collective}, we have incorporated these different factors into a random walk model and explored the extent to which different combinations of guiding factors induce migration in a targeted direction, while maintaining a relatively compact configuration. 

The model includes a simplistic description of intracellular signalling, whereby each cell is associated with a set of evolution equations for the levels of membrane-localised \textit{Rac1} activation that determine the movement probability into different directions; each evolution equation is positively or negatively regulated according to local guidance cues. This level of detail allows certain subtleties to be incorporated, such as, for example, the suggestion that the chemoattractant only stabilizes and reinforces the existing \textit{Rac1} protrusions created by cell-cell contacts in cranial NCC in \textit{Xenopus} \cite{theveneau2010collective}. In the light of such findings, the model setting that includes a Hill-type response to the chemoattractive profile is perhaps more realistic, as it assumes the chemoattractive cue could only be sensed at a membrane location if \textit{Rac1} is sufficiently activated at that point. 

The inclusion of intracellular signalling also allows persistence of polarity, where the cell maintains a preferred direction of motion over long time intervals, to emerge naturally.  First, slowing the rate of Rac1 inactivation will act to maintain a high Rac1 level in a certain direction, potentially long after the initial polarising event. Coupled to an asymmetric bias function that curbs switching the axis of movement direction, persistent straight line movements can arise. Thus, according to the length of memory encoded by Rac1 inactivation, model dynamics can switch between an uncorrelated and correlated random walk. Note that while the asymmetric bias function allows strong bias, it is potentially unsuitable for cases where weak individual migration towards a chemical attractor is observed. The symmetric bias function instead introduces more randomness, and subsequently requires more cell-cell and cell-environment interactions for efficient collective migration.

In scenarios where cell polarity is less persistent, the combination of CIL and co-attraction, along with the Hill-like \textit{Rac1} activation rate by the chemoattractant, performs better than the other scenarios. Co-attraction is the only included cell-cell interaction that can lead to clustering, while CIL is the sole cell-cell interaction that promotes stronger dispersion. Despite this, scenarios in which CIL is included perform better than cases without CIL, suggesting that it might play an important role in promoting robust directed migration. We note that when the chemoattractive cue is relatively strong or, alternatively, when the response rate to the chemoattractant is relatively large, cell-cell interactions become less relevant. Effectively, chemotactic behaviour dominates the dynamics.

Despite our inclusion of some detail at the intracellular level, there are many further aspects that can be considered to refine the model if needed. One potential direction would be to, in addition to the \textit{Rac1} dynamics, also consider the \textit{RhoA} evolution, similar to \cite{merchant2018rho, merchant2020rho}. Another alternative would be to incorporate the degradation of the chemoattractant by cells, similar to \cite{mclennan2012multiscale, mclennan2015neural, mclennan2015vegf}. This mechanism, in combination with the Hill-like switch for sensing the chemoattractant, might lead to emergent leader-follower behavior since only the outer cells of the cluster would be able to sense the chemoattractant, similar to findings in \cite{mclennan2012multiscale, mclennan2015neural, mclennan2015vegf}. If the decision is  dependent on the absolute \textit{Rac1} level, outer cells would become more confident in their direction, and the other cells would follow due to the co-attraction mechanism.

It is noted that a number of other agent-based models have been developed to describe NCC (and other) cell migration processes. Single site lattice models of cellular automaton type (for examples, \cite{binder2012,mort2016}) offer a relatively simple and computationally efficient framework: a cell occupies a lattice site and probabilistically moves (or proliferates) into an adjacent site according to a set of rules based on the neighbourhood state, e.g. occupancy and interactions with the nearest neighbours. A particular advantage of these models lies in an established framework that allows scaling to a continuous partial differential equation for the density distribution, permitting further analysis, e.g. \cite{simpson2007}. A disadvantage lies in their relatively course scale. Other NCC models include finer structure, such as vertex-based approaches where cells form evolving polygonal objects \cite{merchant2018rho} or Cellular Potts Model approaches where a cell occupies a connected set of lattice sites \cite{szabo2016vivo}. These models also have their disadvantages, in particular reliance on simulation instead of analytical insight. To walk a middle path, the model here has assumed that each cell spans 5 lattice sites to form a ``+'' configuration, where the centre represents the cell centre and ODEs are associated to each neighbouring site to describe the internal dynamics. By retaining the underlying framework of a biased random walk on a lattice, the possibility of course-scaling to a continuous model is saved and we have shown how this can be done, at least in principal. Notably, the resulting PDE for the evolving probability distribution for a cell’s position echoes the taxis-type equation obtained for position-jump random walk models in the presence of a chemical signal \cite{stevens1997aggregation}, where the directional bias is proportional to the gradient of a continuous approximation to the internal state variable. The equation for this internal state variable is also represented by a diffusion-advection type equation, where the transport terms for the internal state variable now arise due to the translocation of membrane positions each time a cell moves. 

Further analysis of the continuous problem is certainly a point for future effort, but the intricacy of the derived system (coupled systems of diffusion-advection) demands a dedicated study. In particular, we note that the equation for the internal state variable contains ``self-gradient-following’’ terms that could potentially result in negative diffusion and a loss of regularity. Other points to consider for this study would be the state space for the internal variable. The simplest option, which was performed here, involved collapsing the internal state space variable to the central point of the cell: effectively, the distinct spatial structure contained within the four membrane locations is lost, merged into a single PDE for the internal variable. An alternative and interesting option would be to derive four separate internal state PDEs, each corresponding to a different membrane location. While the model will inevitably become more complex, such an approach could allow the spatial structure at the cellular scale to be retained at the macroscopic or continuous scale. \\

\textbf{Acknowledgments}
VF was supported by the EPSRC Centre for Doctoral Training in Mathematical Modelling, Analysis and Computation (MAC-MIGS) funded by the UK Engineering and Physical Sciences Research Council (grant EP/S023291/1), Heriot Watt University and the University of Edinburgh. KJP is a member of INdAM-GNFM and acknowledges departmental funding through the `MIUR-Dipartimento di Eccellenza' programme. LJS was supported by Chancellor's Fellowship at the University of Edinburgh.\\

\textbf{Author Contributions}
All authors contributed to study conception and design. VF developed the model, implemented and performed numerical simulations, and performed the analysis. The first draft was written by VF and all authors commented on sequential versions of the manuscript. All authors gave final approval for publication and agree to be held accountable for the work performed therein.\\

\textbf{Code availability}
The code for numerical simulations is available upon request. \\

\appendix

\section{Algorithm Layout}
\label{appx:Algorithm}
\begin{algorithm}
\caption{Biased random walk of $N$ cells}\label{algo1}
\begin{algorithmic}[1]
\Require End time $0<T \in \mathbb{N}$
\Require Domain $D = [x_{1, \min},x_{1, \max}(T)] \times [x_{2, \min}, x_{2, \max}] \subset h \mathbb{Z}^2$
\Require Spatial constraints on $h \mathbb{Z}^2 \setminus D$
\Require Chemoattractive profile $S_1(x_1,x_2) \geq 0 $
\Require Initial positions $P^i(0) \in D$ for $i \leq N$
\Require Initial \textit{Rac1} setup
$$
C_i(\xx, 0 ) = \begin{cases} c_0 \quad \text{for} \quad  \xx \in \mathcal{A}(P^i(0)) ,\\
0 \quad \quad  \text{else.} \end{cases}
$$
\For{$t =0, \dots, T$}
    \For{$i=1, \dots, N$}
        \For{$j =1 , \dots N  \vee j \neq i $}
            \If{$\| P^i(t) - P^j(t) \| \leq R$}
                \State Update co-attractive profile $S_2 (\xx, t )$
            \EndIf
            \If{$\mathcal{A}(P^i(t)) \cap \mathcal{A}(P^j(t)) \neq \empty$}
                \State Update repelling profile $S_3^i (\xx, t )$
            \EndIf
        
            \State Update $C_i(\xx,t+1)$ according to ODE
        \EndFor
    \EndFor
     \For{$i=1, \dots, N$}
        \State Update location $P^i(t+1)$ according to updated \textit{Rac1} $C_i(\xx,t+1)$
    \EndFor
\EndFor
\end{algorithmic}
\end{algorithm}

\section{Supplementary result plots}

\begin{figure}
    \centering
    \begin{subfigure}[b]{0.65\textwidth}
    \includegraphics[width=1\textwidth]{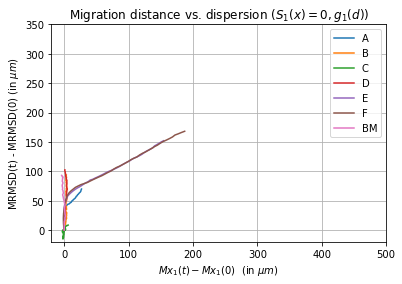}
    \caption{ }
    \label{fig:res:MRMSD_0_g1}
        \end{subfigure} \\
         \begin{subfigure}[b]{0.65\textwidth}
    \includegraphics[width=1\textwidth]{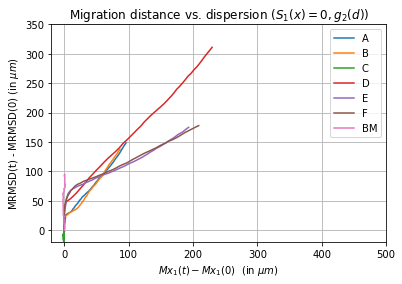}
    \caption{ }
    \label{fig:res:MRMSD_0_g2}
        \end{subfigure} \\   
        \caption{Migration distance $M_{x_1}(t) - M_{x_1}(0)$ vs. dispersion $\text{MRMSD}(t) - \text{MRMSD}(0)$ in the case without a chemoattractant, i.e. $S_1(\xx) = 0$, with the two different gain functions $g_1$ (a) and $g_2$ (b) for $0 \leq t \leq 350 \min$}
\label{fig:Dispersionovertime_S1_0}
\end{figure}

\begin{figure}
     \centering
     \begin{subfigure}[b]{0.49\textwidth}
         \centering
         \includegraphics[width=\textwidth]{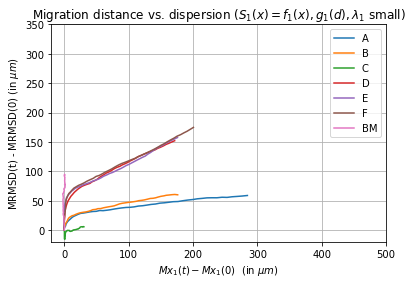}
         \caption{}

     \end{subfigure}
     \begin{subfigure}[b]{0.49\textwidth}
         \centering
         \includegraphics[width=\textwidth]{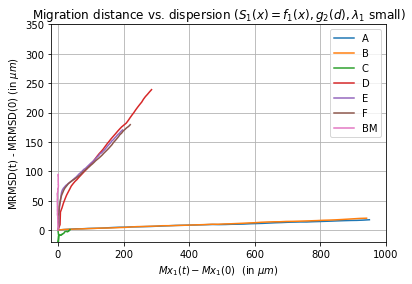}
         \caption{}

     \end{subfigure} \\
     \hfill
     \begin{subfigure}[b]{0.49\textwidth}
         \centering
         \includegraphics[width=\textwidth]{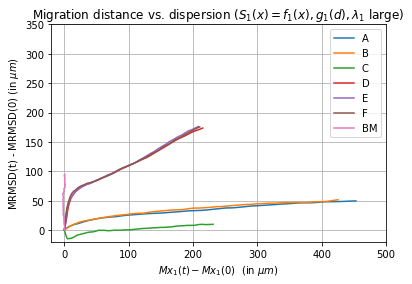}
         \caption{}

     \end{subfigure}
          \begin{subfigure}[b]{0.49\textwidth}
         \centering
         \includegraphics[width=\textwidth]{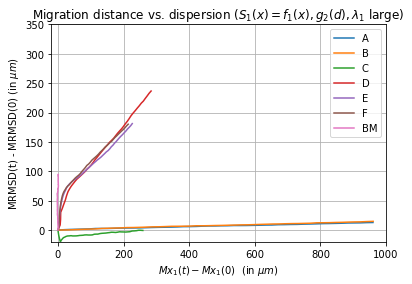}
         \caption{}

     \end{subfigure}
\caption{Mean migration distance vs. dispersion for $S_1 (\xx) = f_1(\xx)$, different bias functions $g_1$ (a,c), $g_2$ (b,d) and activation rates of \textit{Rac1} by the chemoattractant, small $\lambda_1$ (a,b), large $\lambda_1$ (c,d) for $0 \leq t \leq 350 \min$}
\label{fig:Dispersionovertime_S1_f1}
\end{figure}

\begin{figure}
     \centering
     \begin{subfigure}[b]{0.49\textwidth}
         \centering
         \includegraphics[width=\textwidth]{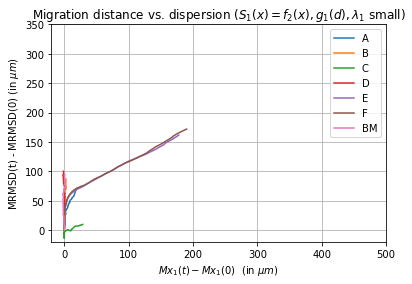}
         \caption{}
         
     \end{subfigure}
     \begin{subfigure}[b]{0.49\textwidth}
         \centering
         \includegraphics[width=\textwidth]{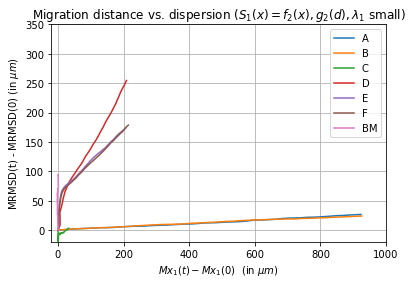}
         \caption{}
        
     \end{subfigure} \\
     \hfill
     \begin{subfigure}[b]{0.49\textwidth}
         \centering
         \includegraphics[width=\textwidth]{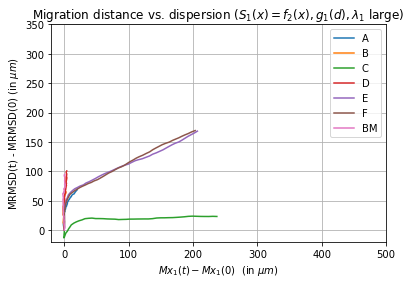}
         \caption{}
         
     \end{subfigure}
          \begin{subfigure}[b]{0.49\textwidth}
         \centering
         \includegraphics[width=\textwidth]{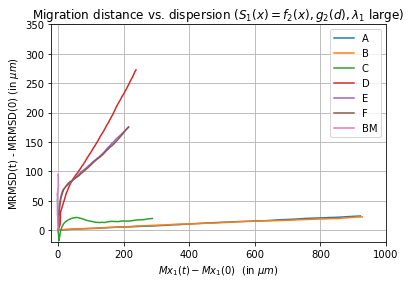}
         \caption{}
        
     \end{subfigure}
        \caption{Mean migration distance vs. dispersion for $S_1 (\xx) = f_2(\xx)$, different bias functions $g_1$ (a,c), $g_2$ (b,d) and activation rates of \textit{Rac1} by the chemoattractant, small $\lambda_1$ (a,b), large $\lambda_1$ (c,d) for $0 \leq t \leq 350 \min$}
        \label{fig:Dispersionovertime_S1_f2}
\end{figure}

\clearpage
\bibliographystyle{plain}
\bibliography{references}

\end{document}